\documentclass[12pt,onecolumn]{IEEEtran}
\usepackage{amssymb}
\usepackage{amsfonts}
\usepackage{amsmath}
\usepackage{algorithm}
\usepackage{graphicx,subfigure,cite,multicol,multirow,diagbox,booktabs,array}

\usepackage[dvips]{color}
\usepackage{float}
\usepackage[noend]{algpseudocode}
\ifCLASSINFOpdf
\else
\fi
\begin{document}
\title{Low-complexity and High-performance Receive Beamforming for Secure Directional Modulation Networks against an Eavesdropping-enabled Full-duplex Attacker}

\author{Yin Teng,~Jiayu Li,~Lin Liu,~Guiyang Xia,~Xiaobo Zhou,~Feng Shu, \\
Jiangzhou Wang,~\emph{Fellow},~\emph{IEEE},~and~Xiaohu You,~\emph{Fellow},~\emph{IEEE}

\thanks{Yin Teng, Jiayu Li, Lin Liu, Guiyang Xia, and Feng Shu are with the School of Electronic and Optical Engineering, Nanjing University of Science and Technology, 210094, CHINA. (e-mail: shufeng0101@163.com). }
\thanks{Xiaobo Zhou is with the School of Physics and Electronic Engineering, Fuyang
Normal University, Fuyang 236037, China. (e-mail:zxb@fynu.edu.cn).}
\thanks{Feng Shu is also with the School of Information and Communication Engineering, Hainan University,~Haikou,~570228, China.}
\thanks{Jiangzhou Wang is with the School of Engineering and Digital Arts, University of Kent, Canterbury CT2 7NT, U.K. (e-mail: j.z.wang@kent.ac.uk).}
\thanks{Xiaohu You is with the School of Information Science and Engineering, Southeast University, 210094, CHINA. (e-mail: xhyu@seu.edu.cn).}

}

\maketitle

%%% Abstract. 摘要
\begin{abstract}
In this paper, we present a novel scenario for directional modulation (DM) networks with a full-duplex (FD) malicious attacker (Mallory), where Mallory can eavesdrop the confidential message from Alice to Bob and simultaneously interfere Bob by sending a jamming signal. Considering that the jamming plus noise at Bob is colored, an enhanced receive beamforming (RBF), whitening-filter-based maximum ratio combining (MRC) (WFMRC), is proposed. Subsequently, two RBFs of maximizing the secrecy rate (Max-SR) and  minimum mean square error (MMSE) are presented to show the same performance as WFMRC. To reduce the computational complexity of conventional MMSE, a low-complexity MMSE is also proposed.  Eventually, to completely remove the jamming signal from Mallory and transform the residual interference plus noise to a white one, a new RBF, null-space projection (NSP) based maximizing WF receive power, called NSP-based Max-WFRP, is also proposed. From simulation results, we find that the proposed Max-SR, WFMRC,   and  low-complexity MMSE have the same SR performance as conventional MMSE, and achieve the best performance while the proposed NSP-based Max-WFRP performs better than MRC in the medium and high signal-to-noise ratio regions. Due to its low-complexity,the proposed low-complexity MMSE is very attractive. More important, the proposed methods are robust to the change in malicious jamming power compared to conventional MRC.
\end{abstract}

%%% Keywords. 关键词
\begin{IEEEkeywords}
Malicious attacker, secure, directional modulation, secrecy rate, receive beamforming, null-space projection (NSP).
\end{IEEEkeywords}

\IEEEpeerreviewmaketitle

\section{Introduction}
In the past decade, physical layer security (PLS) has attracted wide attention from academia \cite{WangDistributed,Chen2017A,ZhaoAnti,WuSecure,Zou2016Relay,Goel2008Guaranteeing,WangSecuring,ShuDirectional}. Directional modulation (DM), as an advanced PLS transmission technique, is suitable for line-of-sight (LoS) propagation channel. Beamforming technology with artificial noise (AN) has a capability of creating a secure transmission in the LoS propagation channel \cite{DingOrthogonal,Feng2017Robust,ShuSecure,WuSecure}.

In \cite{Daly2009Directional}, the authors presented a DM technique that uses the phased arrays to generate modulation and provides security by deliberately distorting signals in other directions and making the signal independent of direction. An orthogonal vector approach was proposed in \cite{DingOrthogonal} for the synthesis of multi-beam DM transmitters. In \cite{Feng2017Robust,HuRobust2016}, the authors developed robust synthesis schemes, in which DM is able to enhance the security performance of desired directions and distort the constellation points of undesired directions. These systems have the capability of concurrently projecting independent data streams along different specified spatial directions while simultaneously distorting signal constellations in all other directions. A directional modulation technique using frequency diverse array (FDA) for secure communications was proposed in \cite{Hu2017Artificial,QiuArtificial,QiuMulti,ChengSVD,ChengWFRFT,XiongDirectional,WangDM}. In \cite{Hu2017Artificial,QiuArtificial}, the authors developed a novel DM scheme based on random frequency diverse arrays with artificial noise (RFDA-DM-AN) to enhance physical layer security of wireless communications. In order to address the limitations of the previous works and further enhance physical layer (PHY) security, an AN-aided index modulation (IM) scheme with cooperative LUs based on FDA beamforming was proposed in \cite{QiuMulti}. Multi-beam directional modulation schemes based on FDA were investigated in \cite{QiuMulti,ChengSVD,ChengWFRFT}. Different from the traditional directional modulation that only implements angle dependent directional modulation, the authors proposed some precise secure transmission schemes of achieving two-dimensional dependence on  angle and range dimensions in \cite{Hu2017Artificial,XiongDirectional,WangDM}. In the multi-carrier based DM antenna array system design, to solve the problems of high peak-to-average-power ratio (PAPR) and phase pattern formation  simultaneously, a method, called wideband beam and phase pattern formation by Newton's (WBPFN),  was proposed in \cite{ZhangMultiCarrier}. In \cite{ZhangBMulti}, the authors developed a multi-carrier based DM framework using antenna arrays, which can realize data transmission on multiple frequencies simultaneously, so as to obtain a higher data rate. A practical wireless transmission scheme was proposed in \cite{ShuSecure} to transmit the confidential message (CM) to the desired user securely and precisely by jointly exploiting multiple techniques, including artificial noise (AN) projection, phase alignment/beamforming, and random subcarrier selection (RSCS) based on orthogonal frequency division multiplexing (OFDM)\cite{ZhuChunk,ZhuChunk2012}, and DM. Particularly, if a DM transmitter intends to transmit CM to Bob and to interfere eavesdropper (Eve) with AN, she needs to know the directions of Bob and Eve in advance. In \cite{ShuLow}, two phase alignment (PA) methods, hybrid analog and digital PA  and hybrid digital and analog PA, were proposed to estimate direction of arrival (DOA) based on the parametric method. In \cite{ZhuangMachine}, the authors proposed an improved hybrid analog-digital (HAD) estimation of signal parameters via rotational invariance techniques (ESPRIT) at HAD transceiver to measure the DOA of a desired user.

However, the aforementioned existing works focused on the scenarios with only a passive eavesdropper Eve having no active malicious attacking capacity. In such a situation, it is hard for Alice to obtain the channel state information (CSI) from Alice to Eve. This is an embarrassing issue usually raised by secure expert. How to address this challenging issue? If Eve behaves like Mallory, with an active malicious attacking ability, this problem naturally disappears. By channel estimating, Alice may obtain the CSI from Alice to Mallory and Bob can obtain the CSI from Mallory to Bob. In \cite{XuDetection}, the authors proposed a jamming detection method for non-coherent single input multiple output (SIMO) systems without CSI. A hybrid wiretapping wireless system with a half-duplex adversary was proposed in \cite{BasciftciOn}, where the adversary can decide to either jam or eavesdrop the transmitter-to-receiver channel. In the presence of a full-duplex eavesdropper having both eavesdropping and jamming capabilities, the authors in \cite{MukA} proposed a novel secure transmission strategy against eavesdropping by the sophisticated adversary. The authors in \cite{ChenSecrecy} studied the secrecy outage probability and mean SR of the system to evaluate the security performance of the system.
In this paper, our focus is on how to suppress the malicious jamming and improve the secure performance by designing receive beamforming (RBF) at Bob in the presence of the malicious jamming.

In this paper, we present a novel DM scenario with a full-duplex (FD) malicious attacker Mallory. Moreover, Mallory  also eavesdrops the CM conveyed from Alice to Bob. To reduce the impact of jamming from Mallory on Bob and improve the secure performance, four RBF methods are proposed. Our main contributions are summarized as follows:

\begin{enumerate}
  \item To alleviate the jamming from Mallory on Bob and improve the secure performance, the conventional MRC RBF at Bob is presented to strengthen the confidential signal. To enhance its performance, a WFMRC RBF is proposed by converting the colored interference plus noise at Bob into a white one with its covariance matrix being a multiple of identity matrix. Then, two RBFs Max-SR and conventional  MMSE are derived.  From simulation results, the proposed two RBFs WFMRC and Max-SR obviously outperform the conventional MRC in the medium and high SNR regions and achieve the same performance as conventional MMSE in terms of SR and BER.
 \item To completely remove the jamming from Mallory, a new RBF of maximizing the WF-based receive power at Bob is proposed to force the malicious jamming onto the null-space (NSP) of the channel spanning from Alice to Bob. By exploiting the rank-one property of channel steering vector, a low-complexity MMSE is proposed to reduce the complexity of conventional MMSE without performance loss.   According to simulation, we find that  the proposed NSP-based Max-WFRP is slightly worse than Max-SR, WFMRC and low-complexity MMSE in terms of the SR performance. Its main advantage is that its SR performance is independent of the change in malicious jamming power. Moreover, the proposed low-complexity  MMSE has achieved  the same  performance as three proposed high-performance RBFs WFMRC, conventional MMSE, and Max-SR with an extremely low-complexity.
 \end{enumerate}

The remainder of this paper is organized as follows. Section II presents the DM system model. In Section III, four schemes of design RBF vector $\mathbf{v}_{BR}$ is proposed and its closed-form expression is given. Simulation and numerical results are shown in Section IV. Finally, we draw our conclusions in Section V.

Notations: Throughout the paper, matrices, vectors, and scalars are denoted by letters of bold upper case, bold lower case, and lower case, respectively. Signs $(\cdot)^T$, $(\cdot)^{-1}$, $(\cdot)^\dag$, $(\cdot)^H$, $\parallel\cdot\parallel$ and $\rm tr(\cdot)$ represent transpose, inverse, Moore-Penrose, conjugate transpose, norm and trace, respectively. $\textbf{I}_N$ denotes the $N\times N$ identity matrix. The notation $\mathbb{E}\{\cdot\}$ represents the expectation operation.

\section{System Model}
\begin{figure}[htp]
\centering
\includegraphics[width=9cm]{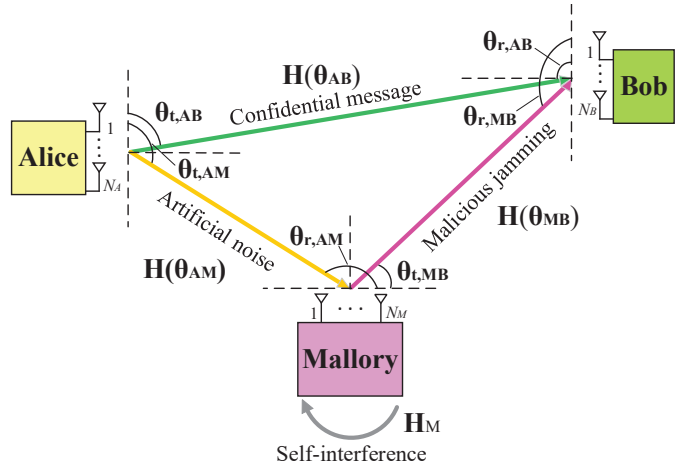}
\caption{Block diagram of the DM network with FD malicious attacker.}
\label{Sys_Mod}
\end{figure}
The proposed DM system is illustrated in Fig.~\ref{Sys_Mod}. It consists of an $N_A$-antennas base station (Alice), a legitimate $N_B$-antennas user (Bob) and an illegal $N_M$-antennas malicious attacker (Mallory). Here, the Alice sends CM to the Bob. In addition, there is an illegal malicious attacker Mallory trying to intercept the CM. Mallory operates on the FD model. That is, he/she can eavesdrop on CM, and simultaneously initiate an active attack on Bob. The baseband transmit signal can be expressed as
\begin{equation}\label{signal sA}
\mathbf{s}_A=\sqrt{\beta_1 P_A}\mathbf{v}_{A}d_A+\sqrt{(1-\beta_1)P_A}\mathbf{T}_{A,AN}\mathbf{z}_{A,AN},
\end{equation}
where $P_A$ is the total transmit power, $\beta_1$ stands for the power allocation (PA) factor, $\mathbf{v}_{A}\in\mathbb{C}^{N_A\times 1}$ denotes the transmit beamforming vector of the CM, and $\mathbf{T}_{A,AN}\in\mathbb{C}^{N_A\times N_A}$ is the
projection matrix for controlling AN to the undesired direction, where $\mathbf{v}^H_{A}\mathbf{v}_{A}=1$ and $\rm tr[\mathbf{T}_{A,AN}\mathbf{T}^H_{A,AN}]=1$. In addition, $\mathbf{v}_A$ is defined as the transmit beamforming vector for the CM. In (\ref{signal sA}), $d_A$ is the CM with average power $\mathbb{E}[\|d_A\|^2]=1$ and $\mathbf{z}_{A,AN}\in\mathbb{C}^{N_A\times1}$ denotes the AN vector with complex Gaussian distribution, i.e., $\mathbf{z}_{A,AN}\sim\mathcal{C}\mathcal{N}(0,\mathbf{I}_{N_A})$.
The malicious attacking signal at Mallory can be expressed as
\begin{equation}\label{signal sm}
\mathbf{s_M}=\sqrt{P_M}\mathbf{T}_{M,AN}\mathbf{z}_{M,AN},
\end{equation}
where $\mathbf{T}_{M,AN}\in\mathbb{C}^{N_{M}\times N_J}$ is the projection matrix for forcing the jamming to Bob, $N_J\in\left\{1,2,....,N_M-1\right\}$. Herein, $P_M$ is the transmit power at Mallory, $\mathbf{z}_{M,AN}\in\mathbb{C}^{N_J\times1}$ denotes the AN vector with complex Gaussian distribution, i.e., $\mathbf{z}_{M,AN}\sim\mathcal{C}\mathcal{N}(0,\mathbf{I}_{N_J})$.

The corresponding received signal at Bob can be written as
\begin{align}\label{Rx_signal rB}
r_B&=\mathbf{v}^H_{BR}(\sqrt{g_{AB}}\mathbf{H}^H(\theta_{AB})\mathbf{s}_A+\sqrt{g_{MB}}\mathbf{H}^H(\theta_{MB})\mathbf{s}_M+\mathbf{n}_B)\nonumber\\
&=\mathbf{v}^H_{BR}(\sqrt{g_{AB}\beta_1 P_A}\mathbf{H}^H(\theta_{AB})\mathbf{v}_{A}d_A +\underbrace{\sqrt{g_{AB}(1-\beta_1)P_A}\mathbf{H}^H(\theta_{AB})\mathbf{T}_{A,AN}\mathbf{z}_{A,AN}}_{\mathbf{n}_A}\nonumber\\
&~~~+\underbrace{\sqrt{g_{MB}}\mathbf{H}^H(\theta_{MB})\mathbf{s}_M}_{\mathbf{n}_M}+\mathbf{n}_B)\nonumber\\
&=\mathbf{v}^H_{BR}(\sqrt{g_{AB}\beta_1 P_A}\mathbf{H}^H(\theta_{AB})\mathbf{v}_{A}d_A+\underbrace{\mathbf{n}_A+\mathbf{n}_M+\mathbf{n}_B}_{\bar{\mathbf{n}}_B}),
\end{align}
where $\mathbf{H}^H(\theta_{AB})=\mathbf{h}(\theta_{r,AB})\mathbf{h}^H(\theta_{t,AB})$ and  $\mathbf{H}^H(\theta_{MB})=\mathbf{h}(\theta_{r,MB})\mathbf{h}^H(\theta_{t,MB})$. $\mathbf{H}^H(\theta_{AB})\in\mathbb{C}^{N_{B}\times N_A}$ denotes the channel matrix from Alice to Bob. $\mathbf{H}^H(\theta_{MB})\in\mathbb{C}^{N_{B}\times N_M}$ denotes the channel matrix from Mallory to Bob.
The normalized steering vector $\mathbf{h}(\theta)$ as
 \begin{align}\label{h_theta}
\mathbf{h}(\theta)=\frac{1}{\sqrt{N}}[e^{j2\pi\Psi_\theta(1)},...,e^{j2\pi\Psi_\theta(n)},...,e^{j2\pi\Psi_\theta(N)}]^T,
\end{align}
where  the phase function $\Psi_\theta(n)$ is defined by \cite{TseFundamentals}
\begin{align}\label{psi_theta}
\Psi_\theta(n)\triangleq-\left(n-\frac{N+1}{2}\right)\frac{d \cos\theta}{\lambda},~~n=~1,\cdots,~N,
\end{align}
where $\theta$ is the  direction of arrival or departure, $n$ is the index of antenna, $d$ represents the element spacing in the transmit antenna array, and $\lambda$ is the wavelength.
In (\ref{Rx_signal rB}), $\mathbf{n}_B\in\mathbb{C}^{N_B\times1}$ is the complex additive white Gaussian noise (AWGN) vector, distributed as $\mathbf{n}_B\sim\mathcal{C}\mathcal{N}(0,\sigma_B^2\mathbf{I}_{N_B})$. Notably, $g_{AB}=\frac{\alpha}{d_{AB}^{c}}$ represents the path loss from Alice to Bob. Here, $d_{AB}$ is the distance between Alice and Bob, $c$ denotes the path loss exponent and $\alpha$ means the path loss at reference distance $d_0$. Moreover, $\mathbf{v}_{BR}\in\mathbb{C}^{N_B\times 1}$ represents the receive beamforming vector of Bob. Using the concept of the null-space projection, we can design
\begin{align}
\mathbf{T}_{A,AN}=\mathbf{I}_{N_A}-\mathbf{H}(\theta_{AB})[\mathbf{H}^H(\theta_{AB})\mathbf{H}(\theta_{AB})]^{-1}\mathbf{H}^H(\theta_{AB}).
\end{align}
Similarly, the receive signal at Mallory is given by
\begin{align}\label{Rx_signal rM}
r_M&=\mathbf{v}^H_{MR}(\sqrt{g_{AM}}\mathbf{H}^H(\theta_{AM})\mathbf{s}_A+\sqrt{\rho}\mathbf{H}^H_{M}\mathbf{s_M}+\mathbf{n}_M)\nonumber\\
&=\mathbf{v}^H_{MR}(\sqrt{g_{AM}\beta_1 P_A}\mathbf{H}^H(\theta_{AM})\mathbf{v}_{A}d_A+\sqrt{g_{AM}(1-\beta_1) P_A}\mathbf{H}^H(\theta_{AM})\mathbf{T}_{A,AN}\mathbf{z}_{A,AN}\nonumber\\
&~~~+\sqrt{\rho P_M}\mathbf{H}^H_{M}\mathbf{T}_{M,AN}\mathbf{z}_{M,AN}+\mathbf{n}_M),
\end{align}
where $\mathbf{H}^H(\theta_{AM})=\mathbf{h}(\theta_{r,AM})\mathbf{h}^H(\theta_{t,AM})$. $\mathbf{H}^H(\theta_{AM})\in\mathbb{C}^{N_{M}\times N_A}$ denotes the channel matrix from Alice to Mallory. $g_{AM}=\frac{\alpha}{d_{AM}^{c}}$ represents the corresponding path loss, and $d_{AM}$is the distance between Alice and Mallory. Additionally, $\mathbf{v}_{MR}$ represents the receive beamforming vector at Mallory. The complex AWGN at Mallory is denoted by $\mathbf{n}_M\sim\mathcal{C}\mathcal{N}(0,\sigma_M^2\mathbf{I}_{N_{M}})$. $\sqrt{\rho} \mathbf{H}^H_{M}\in\mathbb{C}^{N_{M}\times N_{M}}$ is the residual self-interference (RSI) channel matrix of Mallory, $\rho\in[0,1]$ denotes the residual self-interference parameter of the Mallory after self-interference cancelation \cite{MukA}.

As per (\ref{Rx_signal rB}) and (\ref{Rx_signal rM}), we can derive the achievable rate along Bob and Mallory as
\begin{align}\label{RAB}
&R_{AB}(\mathbf{v}_{BR})=\log_2\left(1+\frac{\mathbf{v}^H_{BR}\mathbf{A}\mathbf{v}_{BR}}{ \mathbf{v}^H_{BR}\mathbf{B}\mathbf{v}_{BR}+\mathbf{v}^H_{BR} \mathbf{D}\mathbf{v}_{BR}+\sigma_B^2}\right),
\end{align}
and
\begin{align}\label{RAM}
&R_{AM}=\log_2\left(1+\frac{\mathbf{v}^H_{MR}\mathbf{E}\mathbf{v}_{MR}}{ \mathbf{v}^H_{MR}\mathbf{F}\mathbf{v}_{MR}+\mathbf{v}^H_{MR}\mathbf{R}_{M}\mathbf{v}_{MR}+\sigma_M^2}\right),
\end{align}
respectively, where $ \mathbf{R}_{M}$ is the covariance matrix of RSI channel,
\begin{align}
\mathbf{R}_{M}&= \rho P_M \mathbf{H}^H_{M}\mathbf{T}_{M,AN}\mathbf{T}^H_{M,AN}\mathbf{H}_{M},\\
\mathbf{A}&=g_{AB}\beta_1 P_A \mathbf{H}^H(\theta_{AB})\mathbf{v}_{A}\mathbf{v}^H_{A}\mathbf{H}(\theta_{AB}),\\
\mathbf{B}&=g_{AB}(1-\beta_1) P_A\mathbf{H}^H(\theta_{AB})\mathbf{T}_{A,AN}\mathbf{T}^H_{A,AN}\mathbf{H}(\theta_{AB}),\\
\mathbf{D}&=g_{MB} P_M \mathbf{H}^H(\theta_{MB})\mathbf{T}_{M,AN}\mathbf{T}^H_{M,AN}\mathbf{H}(\theta_{MB}) ,\\
\mathbf{E}&=g_{AM}\beta_1 P_A\mathbf{H}^H(\theta_{AM})\mathbf{v}_{A}\mathbf{v}^H_{A}\mathbf{H}(\theta_{AM}),\\
\mathbf{F}&=g_{AM}(1-\beta_1) P_A \mathbf{H}^H(\theta_{AM})\mathbf{T}_{A,AN}\mathbf{T}^H_{A,AN}\mathbf{H}(\theta_{AM}).
\end{align}
Then we arrive the achievable SR, $R_s$, as follows:
\begin{equation}\label{Rs}
R_s(\mathbf{v}_{BR})=\max\left\{0,R_{AB}(\mathbf{v}_{BR})-R_{AM}\right\}.
\end{equation}

\section{Proposed Four RBF Methods}
In this section, to improve the security rate performance and reduce the effect of jamming from Mallory on Bob, four  RBF schemes are proposed. Additionally, the conventional MRC is also presented as a performance benchmark for the future comparison.
\subsection{Conventional MRC and Proposed WFMRC}
From the definition  of conventional MRC, we have  %The performance improvement is the higher signal-to-noise ratio brought by Array Gain, which in turn leads to better bit errors Rate characteristics., we can directly draw
\begin{align}\label{MRC}
\mathbf{v}^H_{BR}=\frac{\left(\mathbf{H}^H(\theta_{AB})\mathbf{v}_{A}\right)^H}{\|(\mathbf{H}^H(\theta_{AB})\mathbf{v}_{A})^H\|_2},
\end{align}
under the approximation that  $\overline{\mathbf{n}}_B$ is viewed as a white Gaussian noise vector with zero mean and covariance matrix being identity matrix multiplied by a scalar. However, $\overline{\mathbf{n}}_B$ is a  sum of three terms including malicious interference from Mallory, and is colored. Its covariance matrix is
\begin{align}\label{CnB0}
&\mathbf{C}_{\overline{n}_B}= \mathbb{E}\left\{\overline{\mathbf{n}}_B\overline{\mathbf{n}}^H_B\right\}=\mathbf{B}+\mathbf{D}+\sigma_B^2\mathbf{I}_{N_B}
\end{align}
which  is a positive definite Hermitian matrix, i.e., a normal matrix, and has the eigenvalue-decomposition (EVD) form \cite{HornMatrix}
\begin{align}\label{Cnb0}
\mathbf{C}_{\overline{\mathbf{n}}_B}= \mathbf{Q} \Lambda \mathbf{Q}^H,
\end{align}
where $\mathbf{Q}$ is a unitary matrix, $\Lambda$ is a diagonal matrix $\text{diag}~(d_1, ... ,d_{N_B} )$ with
$d_i$ being the i-th eigenvalue of matrix $\mathbf{C}_{\overline{\mathbf{n}}_B}$. We can construct the following WF matrix
\begin{align}\label{Wwf}
\mathbf{W}_{WF}= (\mathbf{Q} \Lambda^{\frac{1}{2}})^{-1}=\Lambda^{-\frac{1}{2}}\mathbf{Q}^H.
\end{align}
Left multiplication of  $\mathbf{r}_B$ by  $\mathbf{W}_{WF}$ leads to
\begin{align}\label{rB1}
\overline{\mathbf{r}}_B&=\mathbf{W}_{WF}\mathbf{r}_B\nonumber\\
&=\sqrt{g_{AB}\beta_1 P_A }\mathbf{W}_{WF}\mathbf{H}^H(\theta_{AB})\mathbf{v}_{A}d_A+\mathbf{W}_{WF}\overline{\mathbf{n}}_B,
\end{align}
where
\begin{align}\label{rB2}
\mathbf{r}_B=\sqrt{g_{AB}\beta_1 P_A}\mathbf{H}^H(\theta_{AB})\mathbf{v}_{A}d_A+\overline{\mathbf{n}}_B.
\end{align}
Obviously, the new noise vector $\mathbf{W}_{WF}\overline{\mathbf{n}}_B$ becomes a white Gaussian vector with convariance matrix being an identity. Now, similar to  (\ref{MRC}), the WFMRC is directly given as
\begin{align}\label{v_brWF}
\overline{\mathbf{v}}^H_{BR}=\frac{(\mathbf{W}_{WF}\mathbf{H}^H(\theta_{AB})\mathbf{v}_{A})^H}{\|(\mathbf{W}_{WF}\mathbf{H}^H(\theta_{AB})\mathbf{v}_{A})^H\|_2}.
\end{align}

\subsection{Proposed Max-SR Method}
Now, we turn to the method of maximizing the SR in (\ref{Rs}). Observing the SR in (\ref{Rs}), it is evident that the available rate at Mallory is independent of $\mathbf{v}_{BR}$. Thus, the optimization problem of the Max-SR
\begin{align}\label{max_SR}
&\max_{\mathbf{v}_{BR}}~~~~R_s(\mathbf{v}_{BR})\nonumber\\
&~\text{s.t.}~~~~~\mathbf{v}_{BR}^H\mathbf{v}_{BR}=1,
\end{align}
reduces to
\begin{align}\label{max_SR1}
&\max_{\mathbf{v}_{BR}}~~~~R_{AB}(\mathbf{v}_{BR})\nonumber\\
&~\text{s.t.}~~~~~\mathbf{v}_{BR}^H\mathbf{v}_{BR}=1,
\end{align}
which can be rewritten as
%\begin{align}\label{max_SR_vbr}
%&\max_{\mathbf{v}_{BR}}~\frac{\mathbf{v}^H_{BR}\mathbf{A}\mathbf{v}_{BR}}{ \mathbf{v}^H_{BR}\mathbf{B}\mathbf{v}_{BR}+ \mathbf{v}^H_{BR} \mathbf{D}\mathbf{v}_{BR}+\sigma_B^2}.
%\end{align}
%Since $R_{AM}$ in Equation (\ref{Rs}) is independent of $\mathbf{v}_{BR}$ and $R_{AB}=\log_2(1+SJNR)$, the above problem is equivalent to maximizing SJNR. In order to facilitate the solution, we simplify the problem in (\ref{max_SR_vbr}) to the following form,
\begin{align}\label{max_SJNR}
&\max_{\mathbf{v}_{BR}}~~~\frac{\mathbf{v}^H_{BR}\mathbf{A}\mathbf{v}_{BR}}{ \mathbf{v}^H_{BR}\mathbf{C}_{\overline{\mathbf{n}}_B}\mathbf{v}_{BR}}\nonumber\\
&~\text{s.t.}~~~~~\mathbf{v}_{BR}^H\mathbf{v}_{BR}=1,
\end{align}
which is the Max-SJNR. Let us define
\begin{align}\label{re_vbr}
 \mathbf{v}^{\prime H}_{BR}=\mathbf{v}^H_{BR}(\mathbf{C}^{\frac{1}{2}}_{\overline{\mathbf{n}}_B})^H,
\end{align}
then SJNR can be expressed as follows
\begin{align}\label{SJNR0}
&\rm SJNR=\frac{\mathbf{v}^{\prime H}_{BR}\overbrace{(\mathbf{C}^{-\frac{1}{2}}_{\overline{\mathbf{n}}_B})^H\mathbf{A}(\mathbf{C}^{-\frac{1}{2}}_{\overline{\mathbf{n}}_B})}^{\mathbf{R}}\mathbf{v}^\prime_{BR}}{ \mathbf{v}^{\prime H}_{BR}\mathbf{v}^\prime_{BR}}~,
\end{align}
Max-SJNR is rewritten as
\begin{align}\label{max_SJNR-Simp}
&\max_{\mathbf{v}^\prime_{BR}}~~~\mathbf{v}^{\prime H}_{BR}\mathbf{R}\mathbf{v}^\prime_{BR}\nonumber\\
&~\text{s.t.}~~~~~\mathbf{v}^{\prime H}_{BR}\mathbf{v}^\prime_{BR}=1.
\end{align}
Let us differentiate the SJNR
expression with respect to $\mathbf{v}^{\prime H}_{BR}$ and set it to zero
\begin{align}\label{d_SJNR}
\frac{\partial}{\partial \mathbf{v}^{\prime}_{BR}}(\frac{\mathbf{v}^{\prime H}_{BR}\mathbf{R} \mathbf{v}^\prime_{BR}}{\mathbf{v}^{\prime H}_{BR}\mathbf{v}^\prime_{BR}})
=\frac{\mathbf{R} \mathbf{v}^\prime_{BR}\cdot\mathbf{v}^{\prime H}_{BR}\mathbf{v}^\prime_{BR}-\mathbf{v}^{\prime H}_{BR}\mathbf{R} \mathbf{v}^\prime_{BR} \cdot\mathbf{v}^\prime_{BR}}{(\mathbf{v}^{\prime H}_{BR}\mathbf{v}^\prime_{BR})^2}=0,
\end{align}
which means that the numerator equals zero and
\begin{align}\label{vbr_MF}
\mathbf{R}\mathbf{v}^\prime_{BR}=\underbrace{\frac{\mathbf{v}^{\prime H}_{BR}\mathbf{R} \mathbf{v}^\prime_{BR}}{\mathbf{v}^{\prime H}_{BR} \mathbf{v}^\prime_{BR}}}_{\lambda}\cdot\underbrace{\mathbf{v}^{\prime}_{BR}}_{\mathbf{v}}.
\end{align}
 Observing the above equation, $\lambda$ is the eigenvalue and $\mathbf{v}$ is the eigenvector associated with $\lambda$. In other words, the maximum SJNR is achieved as $\mathbf{v}^{\prime}_{BR}$ is taken to be the eigenvector corresponding to the largest eigenvalue $\lambda_{max}$ of $\mathbf{R}$. In particular,  the semi-definite positive matrix $\mathbf{R}$  is rank one, written as
\begin{align}\label{decomp-LU}
\mathbf{R}=\mathbf{a}\mathbf{a}^H
\end{align}
with
\begin{align}
\mathbf{a}=\sqrt{g_{AB}\beta_1 P_A}(\mathbf{C}^{-\frac{1}{2}}_{\overline{\mathbf{n}}_B})^H \mathbf{H}^H(\theta_{AB})\mathbf{v}_{A}.
\end{align}
Let us define
\begin{align}\label{eigen-vec}
\mathbf{v}=\frac{\mathbf{a}}{\|\mathbf{a}\|_2},
\end{align}
and multiplying the equality $\mathbf{R}=\mathbf{a}\mathbf{a}^H$ from right by $\mathbf{v}$ yields
\begin{align}\label{vbr_eigemn-equ}
\mathbf{R}\underbrace{\frac{\mathbf{a}}{\|\mathbf{a}\|_2}}_{\mathbf{v}}=\underbrace{\left(\mathbf{a}^H\mathbf{a}\right)}_{\lambda_{max} }\underbrace{\frac{\mathbf{a}}{\|\mathbf{a}\|_2}}_{\mathbf{v}},
\end{align}
which means
\begin{align}
\mathbf{v}=\frac{\mathbf{a}}{\|\mathbf{a}\|_2}
\end{align}
 is the eigenvector corresponding to the largest eigenvalue of matrix $\mathbf{R}$. Clearly,  we have
\begin{align}\label{vbr_CFE}
\mathbf{v}^{\prime\ast}_{BR}=\mathbf{v}_{max}=\mathbf{v}=\frac{\mathbf{a}}{\|\mathbf{a}\|_2},
\end{align}
which is in agreement with the expression in (\ref{v_brWF}).
Substituting the above back into (\ref{re_vbr}), we have
\begin{align}\label{maxsr_vbr}
\mathbf{v}^\ast_{BR}=\mathbf{C}^{-\frac{1}{2}}_{\overline{\mathbf{n}}_B}\mathbf{v}^{\prime\ast}_{BR}=\mathbf{C}^{-\frac{1}{2}}_{\overline{\mathbf{n}}_B}\frac{\mathbf{a}}{\|\mathbf{a}\|_2}.
\end{align}

\subsection{Proposed  low-complexity MMSE Method}
In the following, we propose a low-complexity MMSE algorithm based on the minimum mean square error criterion to design receive beamforming. We can derive the $\mathbf{v}_{BR}$ by the following optimization formula

\begin{align}\label{arg_vbr}
\min_{\mathbf{v}_{BR}}~~~~f(\mathbf{v}_{BR})&=\mathbb{E}\left\{(r_B-d_A)(r_B-d_A)^*\right\}\nonumber\\
&=\rm tr [\mathbf{v}^H_{BR}(g_{AB}\beta_1 P_A  \mathbf{H}^H(\theta_{AB})\mathbf{v}_{A}\mathbf{v}^H_{A}\mathbf{H}(\theta_{AB})+\mathbf{R}_{\mathbf{n}_A}+\mathbf{R}_{\mathbf{n}_M}+\mathbf{R}_{\mathbf{n}_B})\mathbf{v}_{BR}\nonumber\\
&~~~-\sqrt{g_{AB}\beta_1 P_A}\mathbf{v}^H_{BR}\mathbf{H}^H(\theta_{AB})\mathbf{v}_{A}-\sqrt{g_{AB}\beta_1 P_A}\mathbf{v}^H_{A}\mathbf{H}(\theta_{AB})\mathbf{v}_{BR}+1],
\end{align}
where $\mathbf{R}_{\mathbf{n}_A}=\mathbb{E}\left\{\mathbf{n}_A\mathbf{n}^H_A\right\}=\mathbf{B}$, $\mathbf{R}_{\mathbf{n}_M}=\mathbb{E}\left\{\mathbf{n}_M\mathbf{n}^H_M\right\}=\mathbf{D}$ and $\mathbf{R}_{\mathbf{n}_B}=\mathbb{E}\left\{\mathbf{n}_B\mathbf{n}^H_B\right\}=\sigma_B^2\mathbf{I}_{N_B}$.

To obtain the optimal receive beamforming, we need to compute the derivative of $m(\mathbf{v}_{BR})$ with respect to $\mathbf{v}_{BR}$,
\begin{align}\label{mmse_dvbr}
\frac{\partial f(\mathbf{v}_{BR})}{\partial\mathbf{v}_{BR}}&=2(g_{AB}\beta_1 P_A  \mathbf{H}^H(\theta_{AB})\mathbf{v}_{A}\mathbf{v}^H_{A}\mathbf{H}(\theta_{AB})+\mathbf{R}_{\mathbf{n}_A}+\mathbf{R}_{\mathbf{n}_M}+\mathbf{R}_{\mathbf{n}_B})\mathbf{v}_{BR}\nonumber\\ &~~~-2\sqrt{g_{AB}\beta_1 P_A}\mathbf{H}^H(\theta_{AB})\mathbf{v}_{A}.
\end{align}
It is easy to see from Eq.(\ref{mmse_dvbr}) that matrix $g_{AB}\beta_1 P_A  \mathbf{H}^H(\theta_{AB})\mathbf{v}_{A}\mathbf{v}^H_{A}\mathbf{H}(\theta_{AB})+\mathbf{R}_{\mathbf{n}_A}+\mathbf{R}_{\mathbf{n}_M}+\mathbf{R}_{\mathbf{n}_B}$ is positive definite.  Let $\frac{\partial f(\mathbf{v}_{BR})}{\partial\mathbf{v}_{BR}}=0$, we have the conventional MMSE
\begin{align}\label{mmse_vbr1}
\mathbf{v}_{BR}=\sqrt{g_{AB}\beta_1 P_A}(\underbrace{g_{AB}\beta_1 P_A  \mathbf{H}^H(\theta_{AB})\mathbf{v}_{A}\mathbf{v}^H_{A}\mathbf{H}(\theta_{AB})+\mathbf{R}_{\mathbf{n}_A}+\mathbf{R}_{\mathbf{n}_M}+\mathbf{R}_{\mathbf{n}_B}}_{\mathbf{O}})^{-1}\mathbf{H}^H(\theta_{AB})\mathbf{v}_{A}.
\end{align}
 However, the complexity of the conventional MMSE method mentioned above is the cubic function of $N_B$, which will become high as $N_B$ tends to large-scale. Interestingly, since the rank of  matrix $\mathbf{H}^H(\theta_{AB})\mathbf{v}_{A}\mathbf{v}^H_{A}\mathbf{H}(\theta_{AB})$, $\mathbf{R}_{\mathbf{n}_A}$, and $\mathbf{R}_{\mathbf{n}_M}$ in matrix $\mathbf{O}$ are one,  the Sherman-Morrison formula can be applied to provide a low-complexity way of computing the inverse of matrix  $\mathbf{O}$. By exploiting the rank-one property of channel steering vector, a low-complexity MMSE is proposed in what follows.

 By repeatedly making use of the Sherman-Morrison formula
\begin{align}\label{sherman-morrison}
(\mathbf{Z}+\mathbf{u}\mathbf{v}^T)^{-1}=\mathbf{Z}^{-1}-\frac{\mathbf{Z}^{-1}\mathbf{u}\mathbf{v}^T\mathbf{Z}^{-1}}{1+\mathbf{v}^T\mathbf{Z}^{-1}\mathbf{u}}
\end{align}
where $\mathbf{Z}\in\mathbb{R}^{n\times n}$, $\mathbf{u},\mathbf{v}\in\mathbb{R}^{n\times 1}$, $\mathbf{Z}$ and $\mathbf{Z}+\mathbf{u}\mathbf{v}^T$ are invertible,  and $1+\mathbf{v}^T\mathbf{Z}^{-1}\mathbf{u}\neq0$,
we have the low-complexity MMSE method as follows
\begin{align}\label{mmse_vbr}
\mathbf{v}_{BR}&=\sqrt{g_{AB}\beta_1 P_A}\mathbf{O}^{-1}\mathbf{H}^H(\theta_{AB})\mathbf{v}_{A},
\end{align}
where
\begin{align}\label{mmse_O-1}
\mathbf{O}^{-1}
&=\mathbf{K}^{-1}- \frac{g_{MB} P_M \mathbf{K}^{-1} \mathbf{H}^H(\theta_{MB})\mathbf{T}_{M,AN}
 \mathbf{T}^H_{M,AN}\mathbf{H}(\theta_{MB})\mathbf{K}^{-1}}{g_{MB} P_M \mathbf{T}^H_{M,AN}\mathbf{H}(\theta_{MB})\mathbf{K}^{-1}\mathbf{H}^H(\theta_{MB})\mathbf{T}_{M,AN}+1},
\end{align}
where
\begin{align}\label{mmse_K-1}
\mathbf{K}^{-1}&=\mathbf{L}^{-1}-\frac{g_{AB}(1-\beta_1)P_A\mathbf{L}^{-1}\mathbf{H}^H(\theta_{AB})\mathbf{h}(\theta_{t,AB})\mathbf{h}^H(\theta_{r,AB})\mathbf{L}^{-1}}{1+g_{AB}(1-\beta_1)P_A\mathbf{h}^H(\theta_{r,AB})\mathbf{L}^{-1}\mathbf{H}^H(\theta_{AB})\mathbf{h}(\theta_{t,AB})},
\end{align}
where
\begin{align}\label{mmse_L-1}
\mathbf{L}^{-1}&=\mathbf{M}^{-1}+\frac{2g_{AB}(1-\beta_1)P_A\mathbf{M}^{-1}\mathbf{H}^H(\theta_{AB})\mathbf{h}(\theta_{t,AB})\mathbf{h}^H(\theta_{r,AB})\mathbf{M}^{-1}}{1-2g_{AB}(1-\beta_1)P_A\mathbf{h}^H(\theta_{r,AB})\mathbf{M}^{-1}\mathbf{H}^H(\theta_{AB})\mathbf{h}(\theta_{t,AB})},
\end{align}
where
\begin{align}\label{mmse_M-1}
\mathbf{M}^{-1}&=\mathbf{N}^{-1}-\frac{g_{AB}(1-\beta_1) P_A\mathbf{N}^{-1}\mathbf{H}^H(\theta_{AB})\mathbf{h}(\theta_{t,AB})\mathbf{h}^H(\theta_{r,AB})\mathbf{N}^{-1}}{1+g_{AB}(1-\beta_1) P_A\mathbf{h}^H(\theta_{r,AB})\mathbf{N}^{-1}\mathbf{H}^H(\theta_{AB})\mathbf{h}(\theta_{t,AB})},
\end{align}
where
\begin{align}\label{mmse_N-1}
\mathbf{N}^{-1}&=\sigma_B^{-2}\mathbf{I}_{N_B}-\frac{\sigma_B^{-4}g_{AB}\beta_1 P_A\mathbf{H}^H(\theta_{AB})\mathbf{v}_{A}\mathbf{v}^H_{A}\mathbf{H}(\theta_{AB})}{1+g_{AB}\beta_1 P_A\mathbf{v}^H_{A}\mathbf{H}(\theta_{AB})\mathbf{H}^H(\theta_{AB})\mathbf{v}_{A}}.
\end{align}

\emph{Proof}: See Appendix for detailed deriving process.

After we complete the above derivation, the complexity of the proposed low-complexity MMSE  is reduced to the quadratic function of $N_B$.

\subsection{Proposed NSP-based Max-WFRP Method}
We now trun our attention on solving the Max-RP problem, which can be cast as
\begin{align}\label{Max-RP1}
&\max_{\mathbf{v}_{BR}}~~~~\mathbf{v}^H_{BR}\mathbf{H}^H(\theta_{AB})\mathbf{v}_{A}\mathbf{v}^H_{A}\mathbf{H}(\theta_{AB})\mathbf{v}_{BR}\nonumber\\
&~~\text{s.t.}~~~~~(\text{C1})~~~~\mathbf{v}^H_{BR}\mathbf{H}^H(\theta_{MB})=\mathbf{0}_{1\times N_{M}}\nonumber\\
&~~~~~~~~~~~(\text{C2})~~~~\mathbf{v}^H_{BR}\mathbf{v}_{BR}=1,
\end{align}
where constraint (C1) ensures the malicious jamming lying in the null-space of Bob. To simplify the above optimization problem, constraint (C1) implies  $\mathbf{v}_{BR}=\mathbf{G}\widetilde{\mathbf{v}}_{BR}$, and
\begin{align}\label{G}
\mathbf{G}=\mathbf{I}_{N_B}-\mathbf{H}^H(\theta_{MB})[\mathbf{H}(\theta_{MB})\mathbf{H}^H(\theta_{MB})]^{-1}\mathbf{H}(\theta_{MB}),
\end{align}
where $\widetilde{\mathbf{v}}_{BR}$ is defined as a new optimization variable.
Since (C1) has projected Malicious jamming into the null-space of Bob's channel, we obtain the new model,
\begin{align}\label{rB_ubr}
\mathbf{\hat{r}}_B&=\sqrt{g_{AB}\beta_1 P_A}\mathbf{G}^H\mathbf{H}^H(\theta_{AB})\mathbf{v}_{A}d_A+\mathbf{G}^H(\mathbf{n}_A+\mathbf{n}_M+\mathbf{n}_B)\nonumber\\
&=\sqrt{g_{AB}\beta_1 P_A}\mathbf{G}^H\mathbf{H}^H(\theta_{AB})\mathbf{v}_{A}d_A+\mathbf{G}^H(\mathbf{n}_A+\mathbf{n}_B).
\end{align}
Similar to (\ref{rB1}), the left multiplication of  (\ref{rB_ubr}) by $\widetilde{\mathbf{W}}_{WF}$ yields
%\begin{align}\label{rB_wfrp}
%&\widetilde{r}_b=\sqrt{\beta_1 P_s g_{AB}}\widetilde{\mathbf{W}}_{WF}\mathbf{G}^H\mathbf{H}^H(\theta_{AB})\mathbf{v}_{A}d_A\nonumber\\
%&~~~~~~+\widetilde{\mathbf{W}}_{WF}\mathbf{G}^H(\mathbf{n}_A+\mathbf{n}_B),
%\end{align}
%where $\widetilde{\mathbf{W}}_{WF}$ is defined
%\begin{align}\label{Wwf_RP}
%&\widetilde{\mathbf{W}}_{WF}=\left(\mathbb{E}\left\{\mathbf{G}^H(n_A+n_B)(n_A+n_B)^H\mathbf{G}\right\}\right)^{-\frac{1}{2}}\nonumber\\
%&~~~~~~~~=\left(\mathbf{G}^H \mathbf{B}\mathbf{G}+\sigma_B^2\mathbf{G}^H\mathbf{G}\right)^{-\frac{1}{2}}.
%\end{align}
\begin{align}\label{rB_wfrp}
\mathbf{\widetilde{r}}_B&=\sqrt{\beta_1 P_A g_{AB}}\widetilde{\mathbf{W}}_{WF}\mathbf{G}^H\mathbf{H}^H(\theta_{AB})\mathbf{v}_{A}d_A +\widetilde{\mathbf{W}}_{WF}\underbrace{\mathbf{G}^H(\mathbf{n}_A+\mathbf{n}_B)}_{\widetilde{\mathbf{n}}_B},
\end{align}
where $\widetilde{\mathbf{W}}_{WF}$ is defined as
\begin{align}\label{Wwf1}
\widetilde{\mathbf{W}}_{WF}= (\widetilde{\mathbf{Q}} \widetilde{\Lambda}^{\frac{1}{2}})^{-1}=\widetilde{\Lambda}^{-\frac{1}{2}}\widetilde{\mathbf{Q}}^H,
\end{align}
and
\begin{align}\label{CnB01}
\widetilde{\mathbf{Q}}\widetilde{\Lambda}\widetilde{\mathbf{Q}}^H=\mathbf{C}_{\mathbf{\widetilde{n}}_B}= \mathbb{E}\left\{\mathbf{\widetilde{n}}_B\mathbf{\widetilde{n}}^H_B\right\}=\mathbf{G}^H\left(\mathbf{B}+\sigma_B^2\mathbf{I}_{N_B}\right)\mathbf{G}.
\end{align}
As such, the optimization problem of NSP-based Max-WFRP can be written as
\begin{align}\label{Max-RP3}
&\max_{\widetilde{\mathbf{v}}_{BR}}~~~~\widetilde{\mathbf{v}}^H_{BR}\mathbf{J}\widetilde{\mathbf{v}}_{BR}\nonumber\\ &~\text{s.t.}~~~~~~\widetilde{\mathbf{v}}^H_{BR}\widetilde{\mathbf{v}}_{BR}=1,
\end{align}
where
\begin{align}
\mathbf{J}=\widetilde{\mathbf{W}}_{WF}\mathbf{G}^H\mathbf{H}^H(\theta_{AB})\mathbf{v}_{A}\mathbf{v}^H_{A}\mathbf{H}(\theta_{AB})\mathbf{G}\widetilde{\mathbf{W}}^H_{WF}.
\end{align}
Considering that the rank of the semi-definite positive matrix $\mathbf{J}$ is unit,~$\mathbf{J}$ can be written as $\mathbf{J}=\mathbf{b}\mathbf{b}^H$,
with $\mathbf{b}=\widetilde{\mathbf{W}}_{WF}\mathbf{G}^H \mathbf{H}^H(\theta_{AB})\mathbf{v}_{A}$.
Similar to (\ref{max_SJNR-Simp}), the optimal solution of $\widetilde{\mathbf{v}}_{BR}$ is directly given by
\begin{align}\label{ubr_RP}
\widetilde{\mathbf{v}}^\ast_{BR}=\frac{\mathbf{b}}{\|\mathbf{b}\|_2}=\frac{\widetilde{\mathbf{W}}_{WF}\mathbf{G}^H \mathbf{H}^H(\theta_{AB})\mathbf{v}_{A}}{\|\widetilde{\mathbf{W}}_{WF}\mathbf{G}^H \mathbf{H}^H(\theta_{AB})\mathbf{v}_{A}\|_2}.
\end{align}

\subsection{Computational Complexity Analysis}
 In this subsection, we turn to analyze the computational complexities of the above-mentioned schemes. The computational complexities of MRC, WFMRC, Max-SR, NSP-based Max-WFRP, low-complexity MMSE and conventional MMSE are respectively given by
\begin{align}
&C_{MRC}= \mathcal{O}(3N_A N_B+2N_B),\\
&C_{WFMRC}=\mathcal{O}(N^3_B+4N^2_A+7N^2_B+5N_AN_B+3N_BN_M-2N_A-N_B-1),\\
&C_{Max-SR}=\mathcal{O}(N^3_B+4N^2_A+8N^2_B+5N_AN_B+3N_BN_M-2N_A-N_B-1),\\
&C_{Max-WFRP}=\mathcal{O}(4N^3_B+N^3_M+4N^2_A+7N^2_B+4N_AN_B+3N_BN_M+3N^2_M\nonumber\\
&~~~~~~~~~~~~~~~~~~-2N_A-N_M-2),\\
&C_{Low-complexity~MMSE}=\mathcal{O}(36N^2_B+12N_AN_B+6N_BN_M+3N_A+N_M-14N_B-6),\\
&C_{Conventional~MMSE}=\mathcal{O}(N^3_B+2N^2_AN_B+2N^2_BN_A+7N^2_B+N_AN_B+2N_BN_M-N_B-1).
\end{align}
float-point operations (FLOPs). Obviously, their complexities have a decreasing order: NSP-based Max-WFRP$>$Conventional MMSE$>$Max-SR$>$WFMRC$>$Low-complexity MMSE$>$MRC. Observing equations from (56) to (62), we find the fact: as $N_B$ tends to large-scale, the complexities of MRC, low-complexity MMSE , and remaining methods  have the orders $\mathcal{O}(N_B)$,$\mathcal{O}(N^2_B)$, and $\mathcal{O}(N^3_B)$. Clearly, the conventional MRC has the lowest complexity among six methods because its complexity is  a  linear function of $N_B$. The proposed MMSE, Max-SR, and WFMRC have the highest same order complexity $\mathcal{O}(N^3_B)$ FLOPs. The proposed low-complexity MMSE is in between them because its complexity is a quadratic function of $N_B$.
\section{Simulation Results and discussions}
In this section, simulations are done to evaluate the performance of the proposed RBF schemes. System parameters are set as follows: quadrature phase shift keying (QPSK) modulation, $P_A=10W$, $N_A=4$, $N_B=4$, $N_M=4$, $\rho=10^{-11}$, $\beta_1=0.9$, $\theta_{t,AB}=90^{\circ}$, $\theta_{t,AM}=125^{\circ}$, $\theta_{t,MB}=45^{\circ}$, $d_{AB}=1km$, $d_{AM}=4km$, $d_{MB}=3km$, $\sigma_B^2=\sigma_M^2$.

%\begin{figure}[htbp]
%\centering
%\begin{minipage}[t]{0.48\textwidth}
%\centering
%\includegraphics[width=7.8cm,height=6.5cm]{sr_snr_five1.eps}\\
%\caption{Curves of SR versus SNR with $P_M=10W$.}
%\label{SR_SNR_beta}
%\end{minipage}
%\begin{minipage}[t]{0.48\textwidth}
%\centering
%\includegraphics[width=7.8cm,height=6.5cm]{sr_pm_five1.eps}\\
%\caption{Curves of SR versus $P_M$  with $\rm SNR=15dB$ and fixed $P_A$=10W.}
%\label{SR_PM_SNR}
%\end{minipage}
%\end{figure}
\begin{figure}[h]
\centering

\includegraphics[width=8.0cm,height=6.5cm]{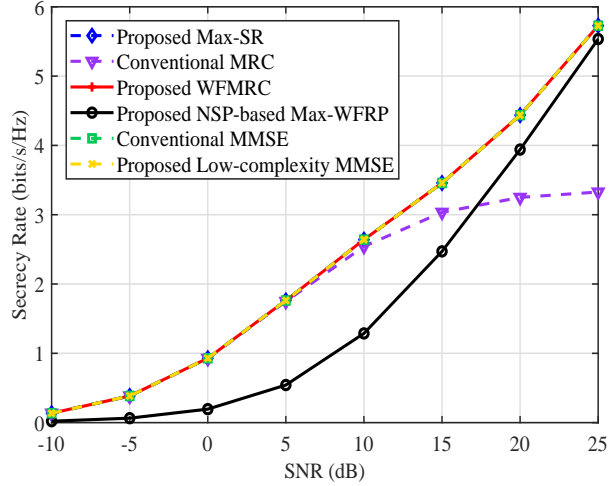}
%\subfigure[$\mathrm{SNR_{new}=5dB}, \mathrm{SNR_{train}=0dB}$]{
%\includegraphics[width=9cm]{5shice-0train_32antennas_2degrees.eps}}
\caption{Curves of SR versus SNR with $P_M=10W$.}
\label{SR_SNR_beta}
\end{figure}

\begin{figure}[h]
\centering

\includegraphics[width=8.0cm,height=6.5cm]{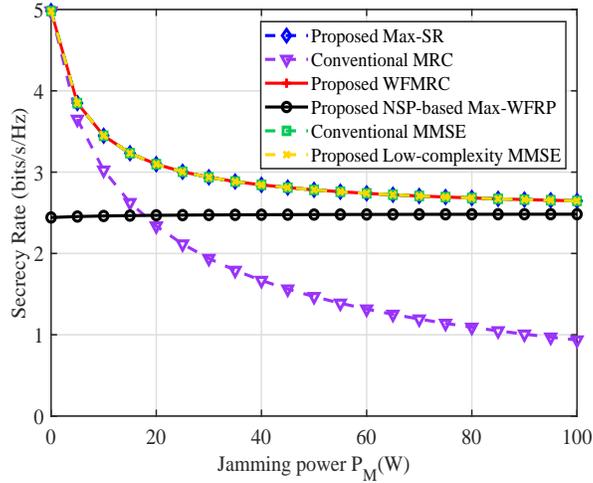}
%\subfigure[$\mathrm{SNR_{new}=5dB}, \mathrm{SNR_{train}=0dB}$]{
%\includegraphics[width=9cm]{5shice-0train_32antennas_2degrees.eps}}
\caption{Curves of SR versus $P_M$  with $\rm SNR=15dB$ and fixed $P_A$=10W.}
\label{SR_PM_SNR}
\end{figure}

Fig.~\ref{SR_SNR_beta} plots the curves of SR  versus SNR of the four proposed methods with $P_M=10$W, where conventional MRC and MMSE are  used as a performance benchmark. It can be seen from this figure that the proposed Max-SR, WFMRC,  and   low-complexity MMSE have the same SR performance as conventional  MMSE, and achieve the best SR performance among all six methods. In the low SNR region, the SR performance of conventional MRC is close to those of the proposed Max-SR, WFMRC,   and low-complexity MMSE and better than that of NSP-based Max-WFRP. The proposed NSP-based Max-WFRP performs much better than MRC and worse than the proposed WFMRC, Max-SR,  and  low-complexity MMSE in the medium and high SNR regions.

Fig.~\ref{SR_PM_SNR} demonstrates the curves of SR versus $P_M$ for the proposed five different methods  with $\rm SNR=15dB$  and fixed $P_A$=10W, where conventional MRC and MMSE are still used as a performance benchmark. From Fig.~\ref{SR_PM_SNR}, it is seen that regardless of the value of $P_M$, the proposed WFMRC, Max-SR,  and   low-complexity MMSE  still perform much better than the proposed NSP-based Max-WFRP and conventional MRC in terms of SR. As $P_M$ increases, the SR performance of the proposed WFMRC, Max-SR , and  low-complexity MMSE  degrade gradually and finally converge to a SR floor. However, this SR floor is still larger than that of NSP-based Max-WFRP with NSP. Interestingly, the SR of NSP-based Max-WFRP with NSP keeps constant and is independent of the change in value of $P_M$ due to the use of NSP operation.  Additionally, the conventional MRC shows a dramatic reduction on SR as $P_M$ increases. This means that the proposed NSP-based Max-WFRP with NSP is the most robust one among six methods. The  conventional MRC  becomes non-robust as the jamming power  $P_M$  increases. The  proposed remaining  methods are in between  NSP-based Max-WFRP  and MRC in accordance with robustiness.

\begin{figure}[h]
\centering

\includegraphics[width=8.0cm,height=6.5cm]{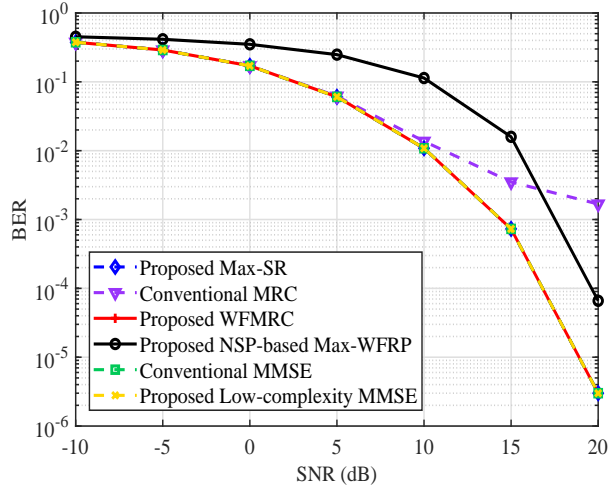}
%\subfigure[$\mathrm{SNR_{new}=5dB}, \mathrm{SNR_{train}=0dB}$]{
%\includegraphics[width=9cm]{5shice-0train_32antennas_2degrees.eps}}
\caption{Curves of BER versus SNR with $P_M$=10W.}
\label{BER_SNR_beta}
\end{figure}

Fig.~\ref{BER_SNR_beta} demonstrates the curves of BER versus SNR of the  proposed five methods with $P_M=10$W, where conventional MRC is still used as a BER performance benchmark.  From Fig.~\ref{BER_SNR_beta}, we can obtain the same performance tendency as Fig.~\ref{SR_SNR_beta}. The proposed Max-SR, WFMRC, and  low-complexity MMSE still have the best performance in our BER simulation. In the low SNR region, the NSP-based Max-WFRP has the worst BER performance, however when the SNR increases up to a certain level, for example SNR=20dB, its performance will be better than that of conventional MRC. Six RBF methods have an increasing order on BER performance: MRC, NSP-based Max-WFRP, Max-SR $\approx$ WFMRC $\approx$  conventional MMSE$\approx$  low-complexity MMSE.

%\begin{figure}[h]
%\centering
%
%\includegraphics[width=8.0cm,height=6.5cm]{ber_snr_five1.eps}
%%\subfigure[$\mathrm{SNR_{new}=5dB}, \mathrm{SNR_{train}=0dB}$]{
%%\includegraphics[width=9cm]{5shice-0train_32antennas_2degrees.eps}}
%\caption{Curves of complexities versus the number of antennas at Bob. }
%\label{BER_SNR_beta}
%\end{figure}

\section{Conclusion}
In this paper, we have investigated RBF schemes in a  DM network with a FD attacker Mallory. Four high-performance RBF schemes, WFMRC, Max-SR, NSP-based Max-WFRP,  and  low-complexity MMSE were proposed to mitigate the impact of the jamming signal on Bob. First, the conventional MRC was presented to  strengthen CM. Due to the colored the jamming plus noise at Bob, the WFMRC was proposed to convert the colored noise to a white one. Then, the Max-SR and low-complexity MMSE were proposed and shown to be equivalent to WFMRC. Eventually, to completely remove the jamming from Mallory, a NSP-based Max-WFRP was  proposed. Moreover, their closed-form expressions were derived with  extremely low-complexities. Simulation results show that the proposed four methods perform much better than the conventional MRC in the medium and high SNR regions, and their SR performances are in an increasing order: MRC, NSP-based Max-WFRP, and Max-SR=WFMRC=low-complexity MMSE. Compared with conventional MRC, the proposed five methods are more robust against the change of jamming power at Mallory. In particular, the proposed NSP-based Max-WFRP is independent of  the change in jamming power at Mallory, and the most robust one among six methods. More importantly, the proposed low-complexity MMSE achieve the same optimal performance as  WFMRC, Max-SR, and  conventional MMSE while its complexity is one order of magnitude lower than those of the latter methods. This is very attractive.

%\section*{Appendix:~ Derivation of low-complexity MMSE}

\appendix[Appendix:~Derivation of low-complexity MMSE]
%It is easy to see from Eq.(\ref{mmse_dvbr}) that matrix $g_{AB}\beta_1 P_A  \mathbf{H}^H(\theta_{AB})\mathbf{v}_{A}\mathbf{v}^H_{A}\mathbf{H}(\theta_{AB})+\mathbf{R}_{\mathbf{n}_A}+\mathbf{R}_{\mathbf{n}_M}+\mathbf{R}_{\mathbf{n}_B}$ is positive definite.  Let $\frac{\partial f(\mathbf{v}_{BR})}{\partial\mathbf{v}_{BR}}=0$, we obtain
%\begin{align}\label{mmse_vbr1}
%\mathbf{v}_{BR}=\sqrt{g_{AB}\beta_1 P_A}(g_{AB}\beta_1 P_A  \mathbf{H}^H(\theta_{AB})\mathbf{v}_{A}\mathbf{v}^H_{A}\mathbf{H}(\theta_{AB})+\mathbf{R}_{\mathbf{n}_A}+\mathbf{R}_{\mathbf{n}_M}+\mathbf{R}_{\mathbf{n}_B})^{-1}\mathbf{H}^H(\theta_{AB})\mathbf{v}_{A}.
%\end{align}
In order to reduce the complexity of the conventional MMSE in (\ref{mmse_vbr1}), let us first define a new matrix
\begin{align}\label{mmse_Oa}
\mathbf{O}&=g_{AB}\beta_1 P_A  \mathbf{H}^H(\theta_{AB})\mathbf{v}_{A}\mathbf{v}^H_{A}\mathbf{H}(\theta_{AB})+\mathbf{R}_{\mathbf{n}_A}+\mathbf{R}_{\mathbf{n}_M}+\mathbf{R}_{\mathbf{n}_B}\nonumber\\
&=g_{AB}\beta_1 P_A\mathbf{H}^H(\theta_{AB})\mathbf{v}_{A}\mathbf{v}^H_{A}\mathbf{H}(\theta_{AB})+g_{AB}(1-\beta_1) P_A\mathbf{H}^H(\theta_{AB})\mathbf{T}_{A,AN}\mathbf{T}^H_{A,AN}\mathbf{H}(\theta_{AB})\nonumber\\
&~~~+g_{MB} P_M \mathbf{H}^H(\theta_{MB})\mathbf{T}_{M,AN}\mathbf{T}^H_{M,AN}\mathbf{H}(\theta_{MB})+\sigma_B^2\mathbf{I}_{N_B},
\end{align}
where the rank of matrix $\mathbf{H}^H(\theta_{AB})\mathbf{v}_{A}\mathbf{v}^H_{A}\mathbf{H}(\theta_{AB})$, $\mathbf{R}_{\mathbf{n}_A}$, and $\mathbf{R}_{\mathbf{n}_M}$ in matrix $\mathbf{O}$ are one.
The receive beamforming at Bob can be rewritten as
\begin{align}\label{mmse_vbr2a}
\mathbf{v}_{BR}=\sqrt{g_{AB}\beta_1 P_A}\mathbf{O}^{-1}\mathbf{H}^H(\theta_{AB})\mathbf{v}_{A}.
\end{align}
In order to compute the inverse of matrix $\mathbf{O}$,   we rearrange its inverse in the following form
\begin{align}\label{mmse_O-1a}
\mathbf{O}^{-1}&=[\underbrace{\sigma_B^2\mathbf{I}_{N_B}+\mathbf{A}+g_{AB}(1-\beta_1) P_A\mathbf{H}^H(\theta_{AB})\mathbf{T}_{A,AN}\mathbf{T}^H_{A,AN}\mathbf{H}(\theta_{AB})}_{\mathbf{K}}\nonumber\\
&~~~+\underbrace{g_{MB} P_M \mathbf{H}^H(\theta_{MB})\mathbf{T}_{M,AN}}_{\mathbf{u}_{\mathbf{O}}}\underbrace{\mathbf{T}^H_{M,AN}\mathbf{H}(\theta_{MB})}_{\mathbf{v}_{\mathbf{O}}^T}]^{-1}\nonumber\\
&=(\mathbf{K}+\mathbf{u}_{\mathbf{O}}\mathbf{v}_{\mathbf{O}}^T)^{-1}\nonumber\\
\end{align}
Observing the above expression, it is clear that $\mathbf{u}_{\mathbf{O}}$ and $\mathbf{v}_{\mathbf{O}}$ are the rank-one column vectors, using the Sherman-Morrison formula, we directly have
\begin{align}\label{mmse_O-1-1a}
\mathbf{O}^{-1}&=\mathbf{K}^{-1}- \frac{\mathbf{K}^{-1} \mathbf{u}_{\mathbf{O}}\mathbf{v}_{\mathbf{O}}^T
 \mathbf{K}^{-1}}{1+\mathbf{v}_{\mathbf{O}}^T\mathbf{K}^{-1}\mathbf{u}_{\mathbf{O}}}\nonumber\\
&=\mathbf{K}^{-1}- \frac{g_{MB} P_M \mathbf{K}^{-1} \mathbf{H}^H(\theta_{MB})\mathbf{T}_{M,AN}
 \mathbf{T}^H_{M,AN}\mathbf{H}(\theta_{MB})\mathbf{K}^{-1}}{1+g_{MB} P_M \mathbf{T}^H_{M,AN}\mathbf{H}(\theta_{MB})\mathbf{K}^{-1}\mathbf{H}^H(\theta_{MB})\mathbf{T}_{M,AN}},
\end{align}
Similarly, in the following, by repeatedly making use of the Sherman-Morrison formula four times, we get the inverse matrix
\begin{align}\label{mmse_K-1a}
\mathbf{K}^{-1}&=[\underbrace{\sigma_B^2\mathbf{I}_{N_B}+\mathbf{A}+g_{AB}(1-\beta_1) P_A\mathbf{H}^H(\theta_{AB})\mathbf{H}(\theta_{AB})-2g_{AB}(1-\beta_1)P_A\mathbf{H}^H(\theta_{AB})\mathbf{H}(\theta_{AB})}_{\mathbf{L}}\nonumber\\
&~~~+\underbrace{g_{AB}(1-\beta_1)P_A\mathbf{H}^H(\theta_{AB})\mathbf{h}(\theta_{t,AB})}_{\mathbf{u}_{\mathbf{K}}}\underbrace{\mathbf{h}^H(\theta_{r,AB})}_{\mathbf{v}_{\mathbf{K}}^T}]^{-1}\nonumber\\
&=(\mathbf{L}+\mathbf{u}_{\mathbf{K}}\mathbf{v}_{\mathbf{K}}^T)^{-1}\nonumber\\
&=\mathbf{L}^{-1}- \frac{\mathbf{L}^{-1} \mathbf{u}_{\mathbf{K}}\mathbf{v}_{\mathbf{K}}^T
 \mathbf{L}^{-1}}{1+\mathbf{v}_{\mathbf{K}}^T\mathbf{L}^{-1}\mathbf{u}_{\mathbf{K}}}\nonumber\\
&=\mathbf{L}^{-1}-\frac{g_{AB}(1-\beta_1)P_A\mathbf{L}^{-1}\mathbf{H}^H(\theta_{AB})\mathbf{h}(\theta_{t,AB})\mathbf{h}^H(\theta_{r,AB})\mathbf{L}^{-1}}{1+g_{AB}(1-\beta_1)P_A\mathbf{h}^H(\theta_{r,AB})\mathbf{L}^{-1}\mathbf{H}^H(\theta_{AB})\mathbf{h}(\theta_{t,AB})},
\end{align}
where $\mathbf{u}_{\mathbf{K}}$ and $\mathbf{v}_{\mathbf{K}}$ are the rank-one column vectors, which yields
\begin{align}\label{mmse_L-1a}
\mathbf{L}^{-1}&=[\underbrace{\sigma_B^2\mathbf{I}_{N_B}+\mathbf{A}+g_{AB}(1-\beta_1) P_A\mathbf{H}^H(\theta_{AB})\mathbf{H}(\theta_{AB})}_{\mathbf{M}}\underbrace{-2g_{AB}(1-\beta_1)P_A\mathbf{H}^H(\theta_{AB})\mathbf{h}(\theta_{t,AB})}_{\mathbf{u}_{\mathbf{L}}}\nonumber\\
&~~~\bullet\underbrace{\mathbf{h}^H(\theta_{r,AB})}_{\mathbf{v}_{\mathbf{L}}^T}]^{-1}\nonumber\\
&=(\mathbf{M}+\mathbf{u}_{\mathbf{L}}\mathbf{v}_{\mathbf{L}}^T)^{-1}\nonumber\\
&=\mathbf{M}^{-1}- \frac{\mathbf{M}^{-1} \mathbf{u}_{\mathbf{L}}\mathbf{v}_{\mathbf{L}}^T
 \mathbf{M}^{-1}}{1+\mathbf{v}_{\mathbf{L}}^T\mathbf{M}^{-1}\mathbf{u}_{\mathbf{L}}}\nonumber\\
&=\mathbf{M}^{-1}+\frac{2g_{AB}(1-\beta_1)P_A\mathbf{M}^{-1}\mathbf{H}^H(\theta_{AB})\mathbf{h}(\theta_{t,AB})\mathbf{h}^H(\theta_{r,AB})\mathbf{M}^{-1}}{1-2g_{AB}(1-\beta_1)P_A\mathbf{h}^H(\theta_{r,AB})\mathbf{M}^{-1}\mathbf{H}^H(\theta_{AB})\mathbf{h}(\theta_{t,AB})},
\end{align}
where $\mathbf{u}_{\mathbf{L}}$ and $\mathbf{v}_{\mathbf{L}}$ are the rank-one column vectors, which yields
\begin{align}\label{mmse_M-1a}
\mathbf{M}^{-1}&=[\underbrace{\sigma_B^2\mathbf{I}_{N_B}+\mathbf{A}}_{\mathbf{N}}+\underbrace{g_{AB}(1-\beta_1) P_A\mathbf{H}^H(\theta_{AB})\mathbf{h}(\theta_{t,AB})}_{\mathbf{u}_{\mathbf{M}}}\underbrace{\mathbf{h}^H(\theta_{r,AB})}_{\mathbf{v}_{\mathbf{M}}^T}]^{-1}\nonumber\\
&=(\mathbf{N}+\mathbf{u}_{\mathbf{M}}\mathbf{v}_{\mathbf{M}}^T)^{-1}\nonumber\\
&=\mathbf{N}^{-1}- \frac{\mathbf{N}^{-1} \mathbf{u}_{\mathbf{M}}\mathbf{v}_{\mathbf{M}}^T
 \mathbf{N}^{-1}}{1+\mathbf{v}_{\mathbf{M}}^T\mathbf{N}^{-1}\mathbf{u}_{\mathbf{M}}}\nonumber\\
&=\mathbf{N}^{-1}-\frac{g_{AB}(1-\beta_1) P_A\mathbf{N}^{-1}\mathbf{H}^H(\theta_{AB})\mathbf{h}(\theta_{t,AB})\mathbf{h}^H(\theta_{r,AB})\mathbf{N}^{-1}}{1+g_{AB}(1-\beta_1) P_A\mathbf{h}^H(\theta_{r,AB})\mathbf{N}^{-1}\mathbf{H}^H(\theta_{AB})\mathbf{h}(\theta_{t,AB})},
\end{align}
where $\mathbf{u}_{\mathbf{M}}$ and $\mathbf{v}_{\mathbf{M}}$ are the rank-one column vectors, which yields
\begin{align}\label{mmse_N-1a}
\mathbf{N}^{-1}&=[\sigma_B^2\mathbf{I}_{N_B}+\underbrace{g_{AB}\beta_1 P_A\mathbf{H}^H(\theta_{AB})\mathbf{v}_{A}}_{\mathbf{u}_{\mathbf{N}}}\underbrace{\mathbf{v}^H_{A}\mathbf{H}(\theta_{AB})}_{\mathbf{v}_{\mathbf{N}}^T}]^{-1}\nonumber\\
&=(\sigma_B^{2}\mathbf{I}_{N_B}+\mathbf{u}_{\mathbf{N}}\mathbf{v}_{\mathbf{N}}^T)^{-1}\nonumber\\
&=\sigma_B^{-2}\mathbf{I}_{N_B}- \frac{\sigma_B^{-4} \mathbf{u}_{\mathbf{N}}\mathbf{v}_{\mathbf{N}}^T
 }{1+\sigma_B^{-2}\mathbf{v}_{\mathbf{N}}^T\mathbf{u}_{\mathbf{N}}}\nonumber\\
&=\sigma_B^{-2}\mathbf{I}_{N_B}-\frac{\sigma_B^{-4}g_{AB}\beta_1 P_A\mathbf{H}^H(\theta_{AB})\mathbf{v}_{A}\mathbf{v}^H_{A}\mathbf{H}(\theta_{AB})}{1+g_{AB}\beta_1 P_A\mathbf{v}^H_{A}\mathbf{H}(\theta_{AB})\mathbf{H}^H(\theta_{AB})\mathbf{v}_{A}},
\end{align}
where $\mathbf{u}_{\mathbf{N}}$ and $\mathbf{v}_{\mathbf{N}}$ are still the rank-one column vectors.

\ifCLASSOPTIONcaptionsoff
  \newpage
\fi

\bibliographystyle{IEEEtran}
\bibliography{cite}

\end{document}